\documentclass[journal,10pt]{IEEEtran}

\usepackage{booktabs}
\usepackage{multirow}
\usepackage{tabularx}
\usepackage{colortbl}
\usepackage[table,xcdraw]{xcolor}
\usepackage{longtable}
\usepackage{float}
\usepackage{adjustbox}
\usepackage{graphicx}
\usepackage{caption}
\usepackage{subcaption} 

\usepackage{lscape}
\usepackage{enumitem}
\usepackage{tcolorbox}
\usepackage{pifont} 
\usepackage{makecell}
\usepackage{placeins}

\usepackage[colorlinks=true, linkcolor=black, citecolor=black, urlcolor=black]{hyperref}

\usepackage{url}

\begin{document}

\title{Towards AI-Driven RANs for 6G and Beyond: Architectural Advancements and Future Horizons}

\author{%
Mathushaharan~Rathakrishnan,
Samiru~Gayan,~\IEEEmembership{Senior Member,~IEEE,}
Rohit~Singh,~\IEEEmembership{Member,~IEEE,}
Amandeep~Kaur,
Hazer~Inaltekin,
Sampath~Edirisinghe,
and~H.~Vincent~Poor,~\IEEEmembership{Life~Fellow,~IEEE}%

\thanks{Mathushaharan Rathakrishnan and Samiru Gayan are with the Department of Electronic and Telecommunication Engineering, University of Moratuwa, Sri Lanka (e-mail: mathushahar333@gmail.com; samirug@uom.lk).}%
\thanks{Rohit Singh is with Dr B R Ambedkar National Institute of Technology Jalandhar, India (e-mail: rohits@nitj.ac.in).}%
\thanks{Amandeep Kaur is with ABV-IIITM Gwalior, India (e-mail: amandeepkaur@iiitm.ac.in).}%
\thanks{Hazer Inaltekin is with the School of Engineering, Macquarie University, North Ryde, NSW 2109, Australia (e-mail: hazer.inaltekin@mq.edu.au).}%
\thanks{Sampath Edirisinghe is with the Department of Computer Engineering, University of Sri Jayewardenepura, Sri Lanka (e-mail: essedirisinghe@sjp.ac.lk).}%
\thanks{H. Vincent Poor is with the Department of Electrical and Computer Engineering, Princeton University, Princeton, NJ 08544, USA (e-mail: poor@princeton.edu).}%
\thanks{Corresponding author: Samiru Gayan (email: samirug@uom.lk).}%
}

\markboth{SUBMITTED TO IEEE COMMUNICATIONS STANDARDS MAGAZINE}
{Shell \MakeLowercase{\textit{et al.}}: Bare Demo of IEEEtran.cls for IEEE Journals}

\maketitle

\begin{abstract}
It is envisioned that 6G networks will be supported by key architectural principles, including intelligence, decentralization, interoperability, and digitalization. With the advances in artificial intelligence (AI) and machine learning (ML), embedding intelligence into the foundation of wireless communication systems is recognized as essential for 6G and beyond. Existing radio access network (RAN) architectures struggle to meet the ever growing demands for flexibility, automation, and adaptability required to build self-evolving and autonomous wireless networks. In this context, this paper explores the transition towards AI-driven RAN (AI-RAN) by developing a novel AI-RAN framework whose performance is evaluated through a practical scenario focused on intelligent orchestration and resource optimization. Besides, the paper reviews the evolution of RAN architectures and sheds light on key enablers of AI-RAN including, digital twin (DTs), intelligent reflecting surfaces (IRSs), large generative AI (GenAI) models, and blockchain (BC). Furthermore, it discusses the deployment challenges of AI-RAN, including technical and regulatory perspectives, and outlines future research directions incorporating technologies such as integrated sensing and communication (ISAC) and agentic AI. 
\end{abstract}


\IEEEpeerreviewmaketitle

\section{Introduction} 
A new era of connectivity is unfolding with the advancement towards 6G networks. Beyond improvements in data rates, 6G aims to support advanced capabilities such as holographic communication \cite{gong2024holographic}, Internet of Senses (IoS) \cite{10731639}, and other applications that were once considered futuristic. However, delivering these visionary services requires more than a simple iteration of current network architectures. Indeed, the complexity and dynamicity of 6G bring core principles of self-built, self-evolving, and self-healing network infrastructures to the forefront. These core principles and challenges reshape the traditional view of Radio Access Networks (RANs) and position Artificial Intelligence (AI) as a foundational component in future RAN design. 

From 1G through 2G and the beginning part of 3G, traditional rule-based systems and simple algorithms were in place alongside Distributed RAN (D-RAN), with a strong emphasis on basic approaches for traffic prediction and network management. In the later part, this evolved into Machine Learning (ML) and Deep Learning (DL)-based algorithms alongside RAN deployment architectures like Centralized RAN (C-RAN), Virtualized RAN (vRAN), and Open RAN (O-RAN), bringing more advanced tasks such as load balancing, interference management, and anomaly detection while improving scalability, multi-vendor interoperability, and openness. Nevertheless, these generations lacked the autonomy, proactive intelligence, and end-to-end orchestration capabilities required to meet future service demands. Unlike its predecessors, 6G focuses on ubiquitous intelligence powered by large-scale AI, ultimately leading to AI-native networks with minimal human intervention. This requires intelligent orchestration and decision-making at the RAN level, leading to end-to-end automation.
\\ \indent Stemming from the above-mentioned context and extending the discussion towards future-oriented advancements, this paper brings together several innovative aspects, summarized as follows:
\begin{figure*}
    \centering
    \includegraphics[width=0.8\textwidth]{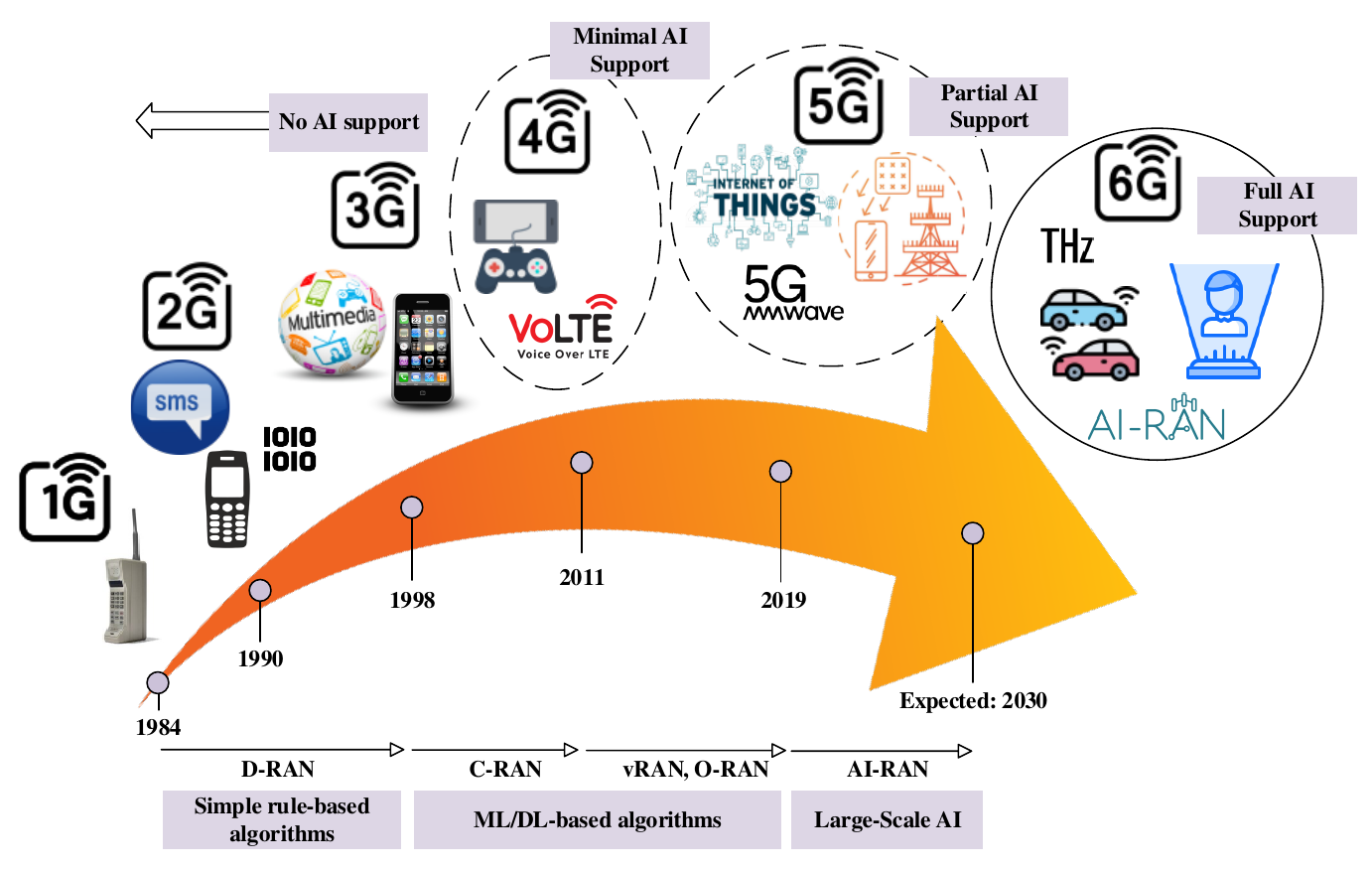} 
    \caption{Correlation among
the evolution of mobile networks and RAN
architectures from the perspective of AI adoption.}
    \label{fig:sample_figure1}
\end{figure*}

\begin{itemize}
   \item This work provides a comprehensive review of the architectural evolution of RANs, emphasizing how pivotal milestones, notably virtualization and openness, have laid the foundation for seamless AI integration. The aim is to offer a novel and forward-looking perspective by reframing the RAN's evolution through the lens of AI readiness, critically analyzing its implications for the development of intelligent and adaptive RAN systems of the future.
   \item Besides, this work systematically identifies and discusses the key enabling technologies driving the evolution of AI-Driven RAN (AI-RAN), with particular emphasis on cutting-edge advancements, highlighting how these emerging innovations collectively pave the way for adaptive, intelligent, and efficient next-generation systems.
   \item Further, this work proposes a proof-of-concept architecture to better align the proposed concepts, validated through extensive simulations for intelligent orchestration and autonomous decision-making. Moreover, the paper analyzes associated challenges in realizing AI-RAN, addressing both technical and regulatory hurdles. Finally, provides an in-depth discussion on emerging research directions and future prospects.
\end{itemize}

\section{Evolution of RANs: Revisiting Architectural Growth}

\begin{figure*}[hbtp]
    \centering
    \includegraphics[width=0.8\textwidth]{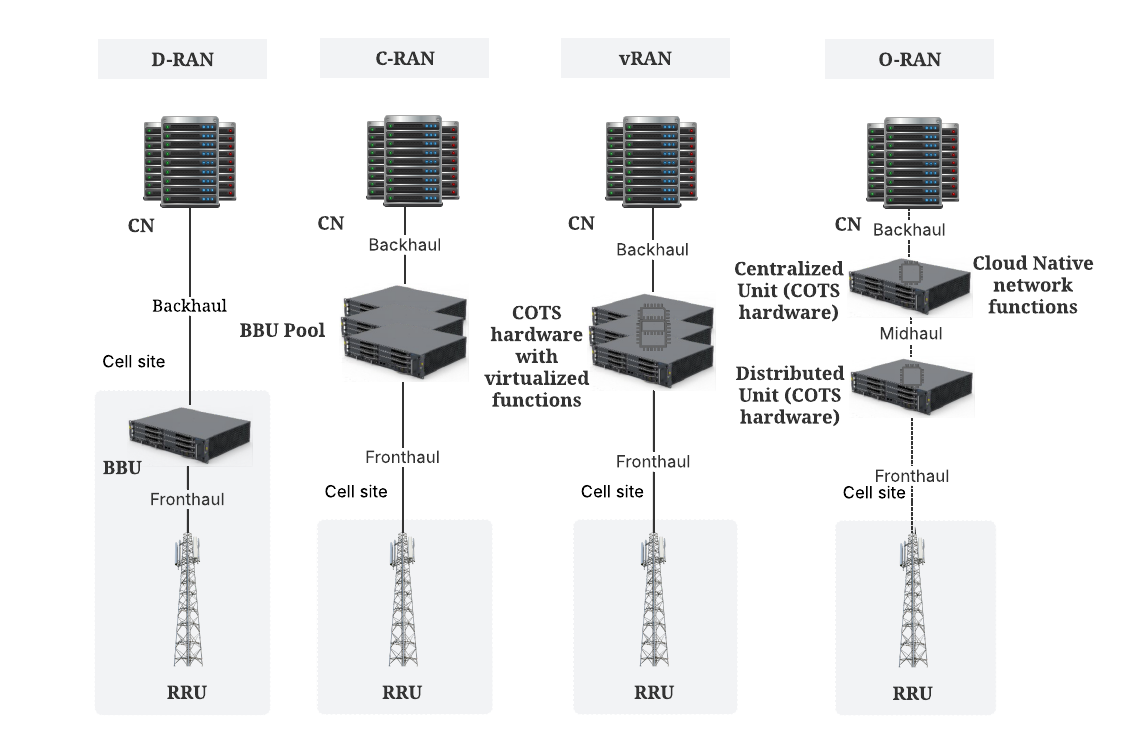} 
    \caption{Architectural transition from D-RAN to O-RAN.}
    \label{fig:sample_figure2}
\end{figure*}

At the heart of wireless systems lies the RAN, the bridge that links User Equipment (UE) with the Core Network (CN). A typical RAN comprises two major components: the Remote Radio Unit (RRU) and the Baseband Unit (BBU). The RRU contains antennas that transmit and receive radio signals, while the BBU orchestrates operations.\footnote{This includes resource utilization, baseband processing (e.g., modulation, demodulation, coding, and decoding), encryption, interference management, and more.} Though early RAN architectures were adequate for the limited scope of wireless systems prior to the Long Term Evolution (LTE) networks, use cases such as Ultra-Reliable Low-Latency Communication (URLLC), Massive Machine-Type Communication (mMTC), and Enhanced Mobile Broadband (eMBB) has necessitated more advanced RAN solutions \cite{frauendorf2023evolution}. Fig.~\ref{fig:sample_figure1} illustrates the correlation between the evolution of mobile networks and the evolution of RAN architectures from the perspective of AI integration in RANs, while Fig.~\ref{fig:sample_figure2} provides a focused view of the architectural transition from D-RAN to O-RAN, which is further elaborated below.

\subsection{D-RAN}
The D-RAN architecture has been a widely used approach, especially from 1G to 3G. In this architecture, the RRU and BBU are co-located at each cell site, facilitating direct on-site processing of radio signals and baseband functions. Each cell site processes the radio signals on the distributed RRU to directly process the radio signals on site, while simultaneously providing the baseband functions on site with the co-located BBU. This setup allows for immediate user data and control signal handling, which provides advantages in terms of reduced latency. With the development of mobile networks, the limitations of D-RAN architecture have become more apparent. Driven by challenges, inefficient resource utilization at all dispersed sites and the inability to apply complex coordination technologies across multiple cell sites, the focus has shifted towards more centralized RAN architectures.

\subsection{C-RAN}
C-RAN offers advantages over D-RAN by centralizing baseband processing into BBU pools, improving throughput, resource efficiency, and power consumption. This centralization reduces both capital and operational costs for service providers. It also enables advanced interference coordination techniques such as Coordinated Multipoint Transmission (CoMP), improving uplink performance by 50\%–100\% \cite{Chih-Lin2014Recent}. Due to these benefits, the C-RAN architecture has been mainly deployed in 4G LTE era and beyond. C-RAN utilizes a fronthaul connection (e.g., Commom Public Radio Interface (CPRI)) to interconnect the centralized BBU pool and RRUs. However, a significant drawback of this architecture is that the interface between the BBU pool and the RRUs remains proprietary.

\subsection{vRAN}

The traditional D-RAN architecture, where all RAN functions are integrated into proprietary hardware, faces limitations in supporting technologies and services required for 5G. vRAN addresses these challenges by virtualizing BBU functions on Commercial Off-The-Shelf (COTS) hardware, enhancing scalability, flexibility, and cost efficiency. It marks the beginning of hardware-software disaggregation in the RAN, eliminating the need for frequent hardware upgrades when introducing new standards or features. This enables flexible scaling and low operational costs. However, like C-RAN, vRAN retains proprietary interfaces, limiting interoperability between vendors and resulting in vendor lock-in. 

\subsection{O-RAN}
With technological advancements, O-RAN, introduced by the O-RAN alliance, has become a promising software-centric architecture that provides network flexibility, cost efficiency, open interfaces, Software-Defined Networking (SDN), multi-vendor capability, Network Function Virtualization (NFV), energy efficiency, and potential advantages for AI integration into RAN \cite{article1}. Beyond its role as an architectural framework, O-RAN acts as a critical enabler for the next evolutionary stage of RAN due to its innate openness, interoperability, and  software-defined capabilities.

\subsubsection*{\textbf{Key Takeaways}}
\textit{It can be inferred from the above discussion that the RAN architecture is undergoing a fundamental transformation toward more openness, intelligence, and automation. Besides, it can be observed that the earlier structures, i.e., D-RAN, C-RAN, and vRAN, though in general provide enhanced efficiency, remained constrained by proprietary interfaces and limited flexibility. Moreover, the advent of O-RAN has addressed these limitations, leveraging the new era of interoperability, programmability, and vendor-neutral design. Further, as we look toward the 6G era, the focus intensifies on creating self-optimizing and fully autonomous network systems. Realizing this ambitious vision demands the integration of AI into the core fabric of RANs, a critical advancement explored in the subsequent sections.}

\section{Key Technology Enablers}
6G is expected to extend existing use cases into Ubiquitous Mobile Ultra-Broadband (uMUB), Ultra-High Data Density (uHDD), and Ultra-High-Speed with Low-Latency Communications (uHSLLC). Apart from already revealed features like openness and intelligence, AI-RAN has the capacity to forecast traffic patterns in the network, dynamically regulate the spectrum, and minimize the necessity of over-provisioning. Furthermore, AI-RAN rectifies network issues and vulnerabilities, thereby improving overall performance and ensuring a future-proof network. Ultimately, AI-RAN aspires to enable networks that are self-adaptive, self-optimizing, and self-sustaining, driven by intelligent and autonomous decision-making processes. This section provides a neutral view of such enhancements driven by the key enabling technologies of AI-RAN.

\subsection{Digital Twin (DT)}
A Digital Twin (DT) is a virtual representation of the components and dynamics of a physical system, and DTs can be broadly categorized into three types: monitoring DTs, simulation DTs, and operational DTs \cite{9711524}. Monitoring DTs support monitoring of the status of a physical system, whereas simulation DTs use various simulation tools to provide future insights. One of the important abilities of AI-RAN is to be self-healing and zero-touch, hence autonomously identify faults and perform live-action responses to ensure self-maintainability. DTs can facilitate automated network maintenance and help reduce management overheads. Another important aspect is that DTs can facilitate AI model/agent training by synthesizing data that represents various network scenarios. Rarely experienced network scenarios can be artificially simulated in a controlled environment to observe network behaviors and assist in the data collection process. By doing so, it removes the burden of collecting large amounts of datasets with controllable distributions to ensure strict Quality of Service (QoS) requirements. Finally, DTs can also improve context awareness by processing multi-sensory data to identify contextual elements related to the environment, time, and situation.  

\subsection{Large Generative AI (GenAI) Models}
The emergence of GenAI marks another inflection point in realizing the vision of future wireless generations, where networks have the capability to adjust, optimize, and reconfigure their functionalities and parameters based on network conditions. This stands in contrast to most of the currently prevailing AI solutions, which are designed for dedicated problem-solving and function merely as add-on features in wireless networks. These are models trained on vast amounts of data to generate new content such as text, images, videos, and more. More specifically, Large Language Models (LLMs) \cite{10685369}, the foundation model for natural language understanding, and Large Action Models (LAMs) \cite{zhang2024xlamfamilylargeaction}, the foundation of AI agents, are the major enablers. Given below are several ways large GenAI models facilitate the development of AI-RAN.

\begin{enumerate}[label=\roman*.]
    \item \textbf{Super-Resolution Localization}: Enhances node positioning by fusing radio signals, spatial data, and environmental context, supporting applications such as Augmented Reality (AR), Virtual Reality (VR), and autonomous mobility.
    \item \textbf{Context-Aware Multi-modal Beamforming}: Predicts optimal beam configurations using multimodal data, improving beam alignment and signal quality while reducing interference and overhead.
    \item \textbf{Enhanced Frequency Division Duplexing (FDD) Transmission}: Infers downlink Channel State Information (CSI) from partial uplink data using attention mechanisms, lowering latency and resource usage in dynamic environments.
\end{enumerate}

\begin{figure*}[htbp]
    \centering
    \begin{subfigure}{0.9\textwidth}
        \centering
        \includegraphics[width=\textwidth]{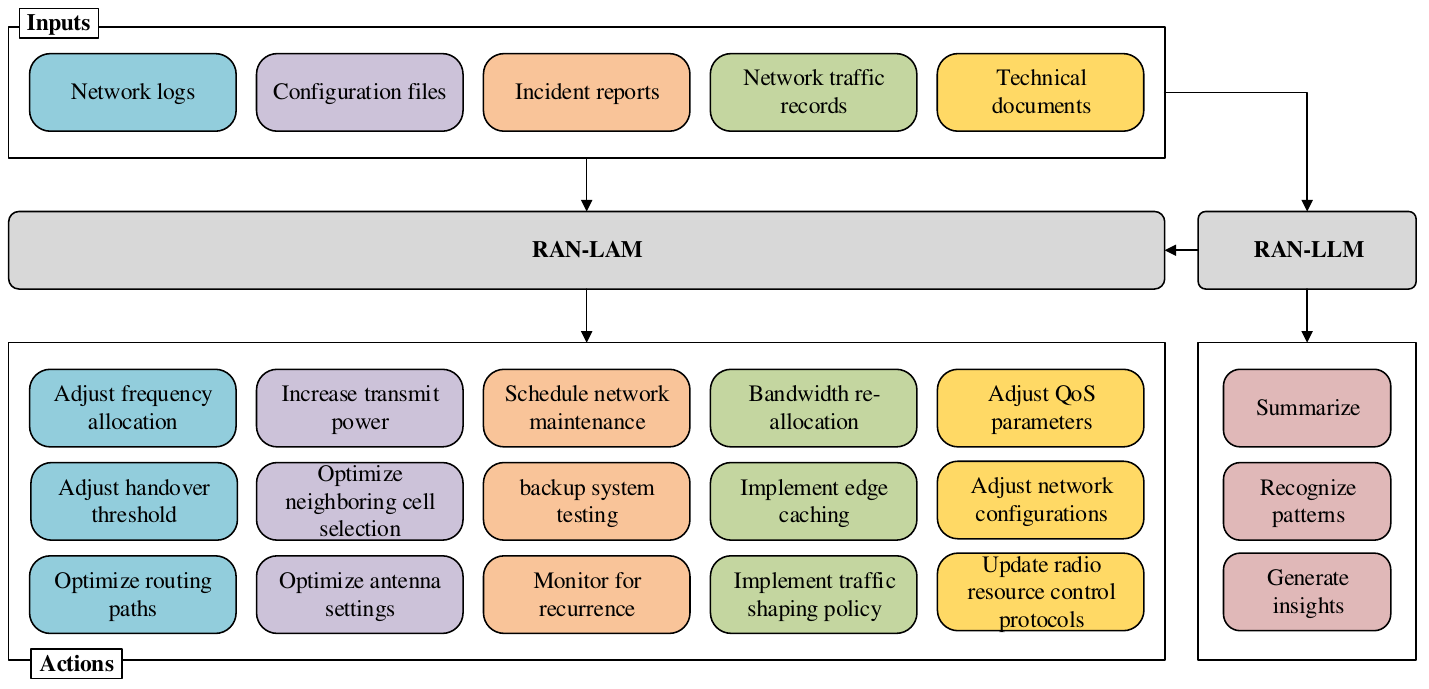}
        \caption{RAN-LAM framework}
        \label{fig:subfig1}
    \end{subfigure}
    
    \vspace{0.5cm}
    
    \begin{subfigure}{0.9\textwidth}
        \centering
        \includegraphics[width=\textwidth]{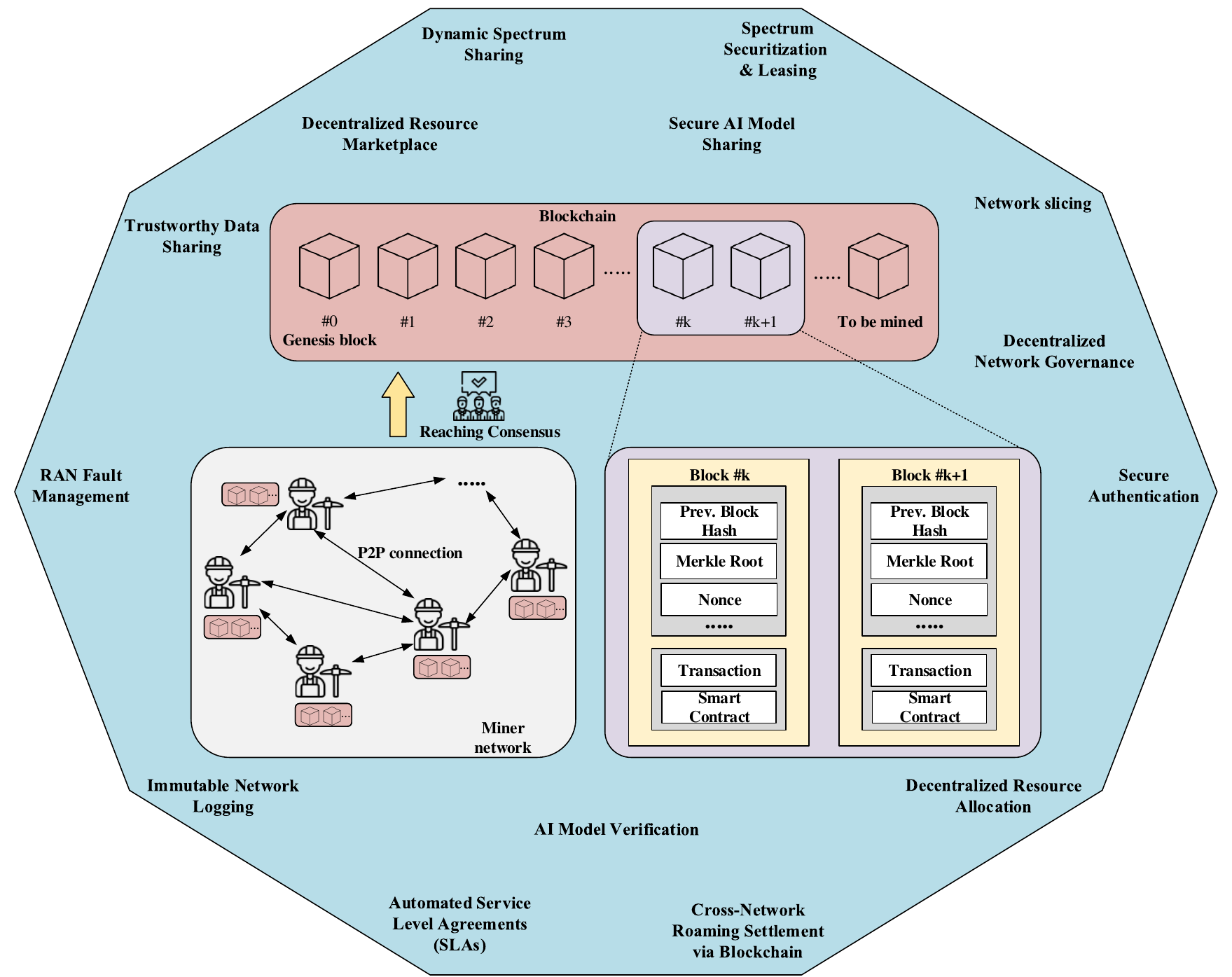}
        \caption{Blockchain workflow and use cases}
        \label{fig:subfig2}
    \end{subfigure}
    
    \caption{Illustration of the RAN-LAM framework and blockchain workflow/use cases for AI-RAN.}
    \label{fig:combined_fig}
\end{figure*}

\subsection{Blockchain (BC)}
The concept of Blockchain (BC), powered by cryptographic hashing, consensus mechanisms, and Merkle trees, gained popularity through the advent of Bitcoin \cite{nakamoto2008} in 2009, where it was introduced as a decentralized ledger to enable secure and tamper-resistant financial transactions without the need for an intermediary. Due to its key intrinsic features, such as transparency, immutability, and decentralization, it has expanded beyond financial applications. It can be used, for example, to deal with network security issues, to prevent roaming fraud, to ensure secure and verifiable Service-Level Agreement (SLA) enforcement, to enable dynamic spectrum and resource sharing, and so on and so forth. Fig. \ref{fig:subfig2} illustrates the fundamental workflow of BC along with use cases.

\subsection{Intelligent Reflecting Surfaces (IRSs)}
Technological advancements that rely primarily on device-side approaches often reach a saturation point due to the unpredictable and uncontrollable nature of the wireless environment. In order to reach the full potential of AI-RAN, it is essential to manipulate the propagation of electromagnetic waves in wireless links, as factors like scattering-induced multipath fading and Doppler effect are detrimental to communication efficiency \cite{9178307}. Intelligent Reflecting Surfaces (IRSs) offer a promising solution by enabling intelligent control over wireless propagation. IRSs are composed of meta-atoms, which are artificially engineered, sub-wavelength structures arranged in a two-dimensional thin layer and designed to exhibit tunable electromagnetic properties with virtually no limits on operating frequencies. Here are some important functionalities of IRSs.

\begin{enumerate}[label=\roman*.]
    \item \textbf{Beam Steering}: Changing the path of an impinging signal to a desired direction by manipulating the reflection or refraction index.
    \item \textbf{Beam Splitting}: Breaking an impinging signal into multiple signals directed towards different orthogonal directions to serve multiple users.
    \item \textbf{Wave Polarization Control}: Controlling the polarization of impinging signals by manipulating their oscillation orientation to achieve customized signal behavior.
    \item \textbf{Wave Absorption}: Minimizing the signal strength of impinging signals to prevent unauthorized access.
    \item \textbf{Localization and Sensing}: Facilitating precise positioning of users and devices in the network by manipulating signals.
    \item \textbf{Simultaneous Wireless Information and Power Transfer (SWIPT)}: IRSs allow signals to carry information and energy simultaneously, facilitating and enabling self-sustaining nodes in AI-RAN.
\end{enumerate}

Physical layer technologies in 6G, such as Ultra-Massive Multiple-Input Multiple-Output (UM-MIMO) \cite{10199530} and Extremely Large-Scale MIMO (XL-MIMO) \cite{10379539}, face challenges due to the need for many Radio Frequency (RF) chains, complex design, high power consumption, and hardware costs, potentially affecting AI-RAN scalability.

\subsection{Federated Learning (FL)}
AI-RAN inherently facilitates rapid growth in ML technologies and their diverse applications. However, a significant bottleneck prevails for many ML-driven use cases due to the requirement of centralized data aggregation from multiple edge devices. To this end, Federated Learning (FL) provides a decentralized approach to ML, where training data remains on edge devices while collaborative training is achieved by sharing model updates and parameters to produce a global model without exposing sensitive training data \cite{10786352}. Table \ref{tab:double-column-example} summarizes some important aspects where FL can make technical impacts in AI-RAN.

\begin{table*}[htbp]
    \caption{Overview of FL contributions for enabling AI-RAN functionalities across key challenges}
    \centering
    \renewcommand{\arraystretch}{1.2}
    \begin{tabularx}{\textwidth}[t]{>{\hsize=0.8\hsize}X 
                                   >{\hsize=1.1\hsize}X 
                                   >{\hsize=1.1\hsize}X 
                                   >{\hsize=1.0\hsize}X}
        \toprule
        \textbf{Aspect} & \textbf{Challenges} & \textbf{Contribution of FL} & \textbf{Key Benefit} \\
        \midrule
        Data privacy &
        Protecting sensitive user data from being transmitted. &
        FL avoids raw data sharing; only updates of the model are exchanged. &
        Reduced attack surface. \\
        
        Real-time decision making &
        Making adaptive, intelligent decisions in real-time with ultra-low latency for mission-critical applications. &
        Enables training and deployment of AI models at the network edge. By processing data locally, FL reduces backhaul transmission. &
        Sub-millisecond decision making. \\
        
        Anomaly detection &
        Vulnerability to cyberattacks, including jamming, Distributed Denial of Service (DDoS), and eavesdropping in AI-RAN. &
        FL-based models collaboratively learn from distributed sources to detect anomalies in real-time. &
        Enhanced AI-RAN security. \\
        
        Beamforming optimization &
        Real-time beamforming in UM-MIMO systems requires accurate CSI for optimal performance. &
        FL accumulates local CSI updates to train global models without raw data sharing, ensuring privacy. &
        Improved spectral efficiency, reduced overhead. \\
        
        Scalability &
        Addressing challenges due to massive connectivity and heterogeneous infrastructure (e.g., macro cells, Unmanned Aerial Vehicles (UAVs)). &
        FL adapts via asynchronous training and model compression, supporting resource-limited nodes. &
        Improved scalability. \\
        \bottomrule
    \end{tabularx}
    \label{tab:double-column-example}
\end{table*}

\subsubsection*{\textbf{Key Takeaways}}
\textit{Emerging technologies, summarized above, hold great potential and emerge to play a crucial role in shaping intelligent and autonomous RAN systems. For instance, DTs, supported by intelligent storage and processing services, enable real-time detection, identification of temporal deviations, and proactive decision-making. Nonetheless, these solutions cannot be directly applied to the emerging applications. For example, the large genAI models, while highly effective in understanding complex patterns and generating insights, face limitations in translating these insights into direct actions or interacting with physical and digital environments. Though emerging solutions exist there, e.g., multiple agents equipped with lightweight AI models can collaboratively use consensus algorithms for network-wide decision-making, as illustrated in Fig.~\ref{fig:subfig1}, where the RAN-LAM framework leverages diverse inputs such as logs, configuration files, and incident reports to optimize performance. Furthermore, emerging techniques like IRSs introduce innovative transceiver designs, allowing signal modulation through IRS elements using a single RF chain, significantly reducing hardware complexity compared to traditional systems where each antenna requires a dedicated RF chain.}

\section{Proof of Concept: Perspective Architecture and Simulation Results}
\begin{figure}
    \centering
    \includegraphics[width=\linewidth]{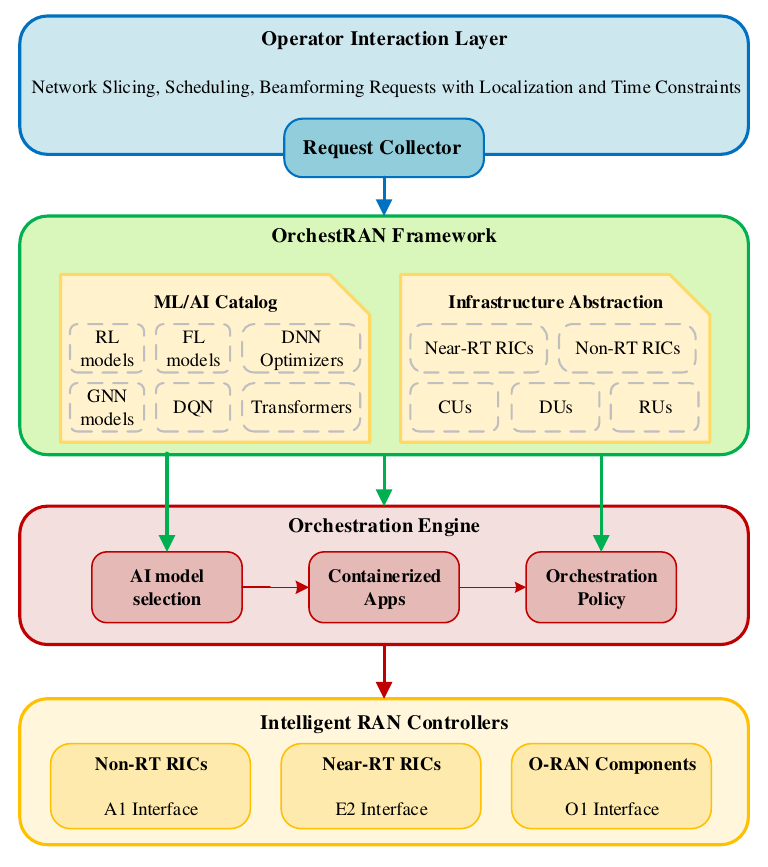} 
    \caption{Proposed AI-RAN architecture for 6G.}
    \label{fig:sample_figure2s}
\end{figure}

We propose an architecture that orchestrates intelligent and adaptive RAN functionalities for 6G networks. As illustrated in Fig.~\ref{fig:sample_figure2s}, it incorporates Orchestration RAN (OrchestRAN) with AI-enhanced optimization and dynamic resource allocation for both Real-Time (RT) and Near-Real-Time (near-RT) applications. Requests from network operators are collected by the Request Collector module, covering functionalities such as network slicing, scheduling, and beamforming, along with associated location (e.g., RAN Intelligent Controller (RIC), Central Unit (CU), and Distributed Unit (DU)) and time constraints (e.g., latency requirements for different service classes). The module aggregates and forwards these to the OrchestRAN framework for processing.

Within OrchestRAN, pre-trained AI models are stored in an ML/AI catalog. This includes Reinforcement Learning (RL) for resource allocation, FL for distributed intelligence, DL optimizers, Graph Neural Networks (GNNs) for topology-aware management, and transformer-based models for traffic forecasting.

The Infrastructure Abstraction module provides a high-level view of physical infrastructure, divided into five logical groups: near-RT RICs, non-RT RICs, CUs, DUs, and RUs. Each group contains a variable number of nodes deployed across the network with virtualization capabilities.

Upon receiving requests, the Orchestration Engine selects suitable AI models from the ML/AI catalog based on service requirements, available computational resources, and current network conditions. These models are embedded into docker containers as dApps, xApps, or rApps, and dispatched to selected nodes. The engine also generates orchestration policies that define data collection for model training and feedback loops for continuous learning. These are converted into executable O-RAN applications using E2, A1, and O1 interfaces.

Intelligent RAN Controllers manage AI-driven control. The non-RT RIC oversees slicing policies and cross-slice optimization, while the near-RT RIC enhances scheduling and beam tracking, manages interfaces, and supports intelligent handovers.

To evaluate performance, we have developed a simulation scenario in Python. The simulation considers an OrchestRAN-based distributed architecture designed to optimize beamforming and scheduling decisions using multi-agent RL.  The scenario includes 100 operator requests per time slot, with 2 non-RT RICs, 5 near-RT RICs, 3 CUs, 8 DUs, and 25 RUs at 28 GHz with 400 MHz bandwidth. The model follows the 3rd Generation Partnership Project (3GPP) TR 38.901 Urban Macro (UMa) path loss model and includes ray-tracing-based fading for both Line-of-Sight (LOS) and Non-Line-of-Sight (NLOS) conditions.

\begin{figure}[htbp]
    \centering
    \begin{subfigure}[b]{0.4\textwidth}
        \centering
        \includegraphics[width=\linewidth]{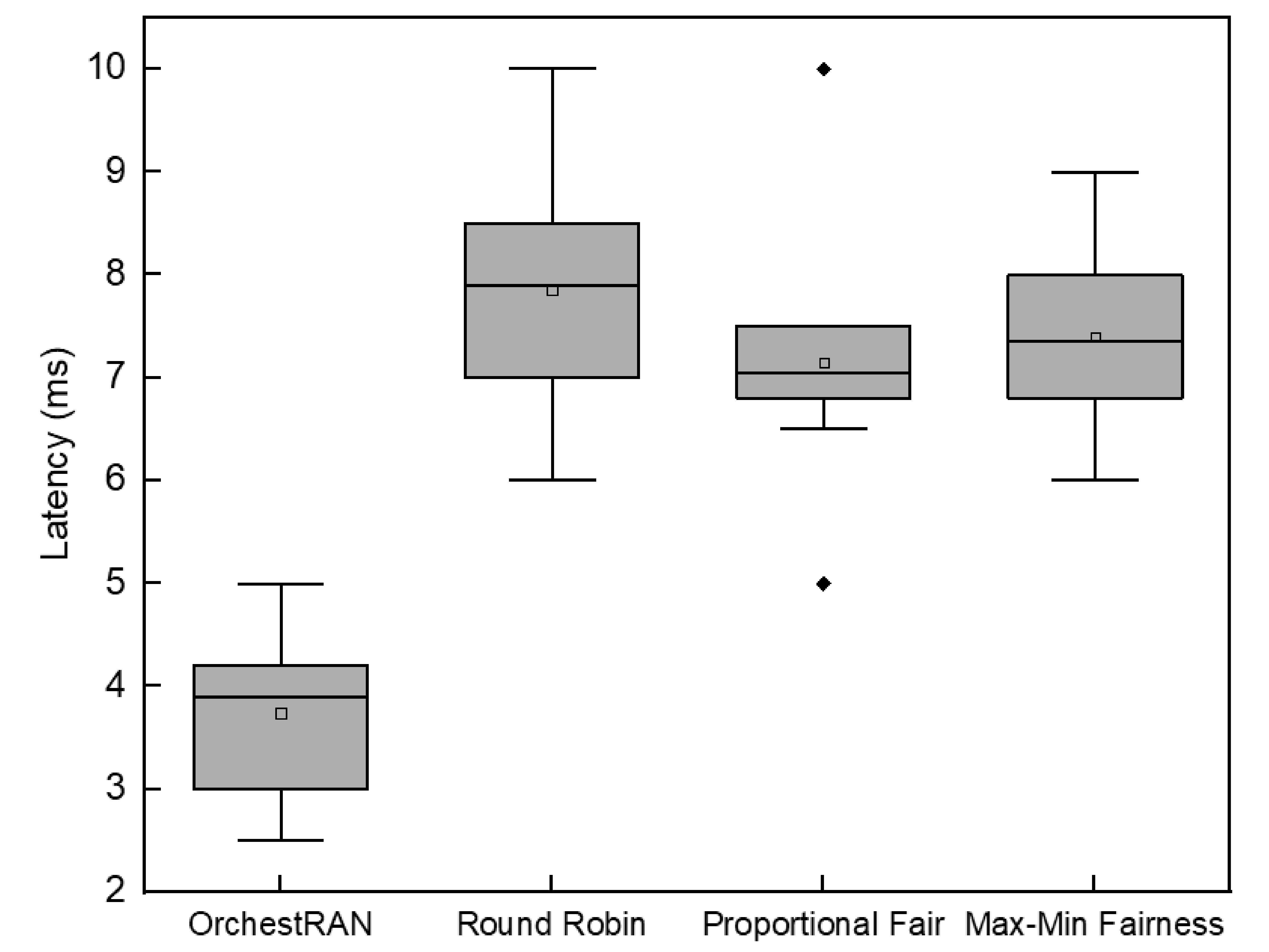}
        \caption{Latency comparison}
        \label{fig:subfig11}
    \end{subfigure}
    \hfill
    \begin{subfigure}[b]{0.45\textwidth}
        \centering
        \includegraphics[width=\linewidth]{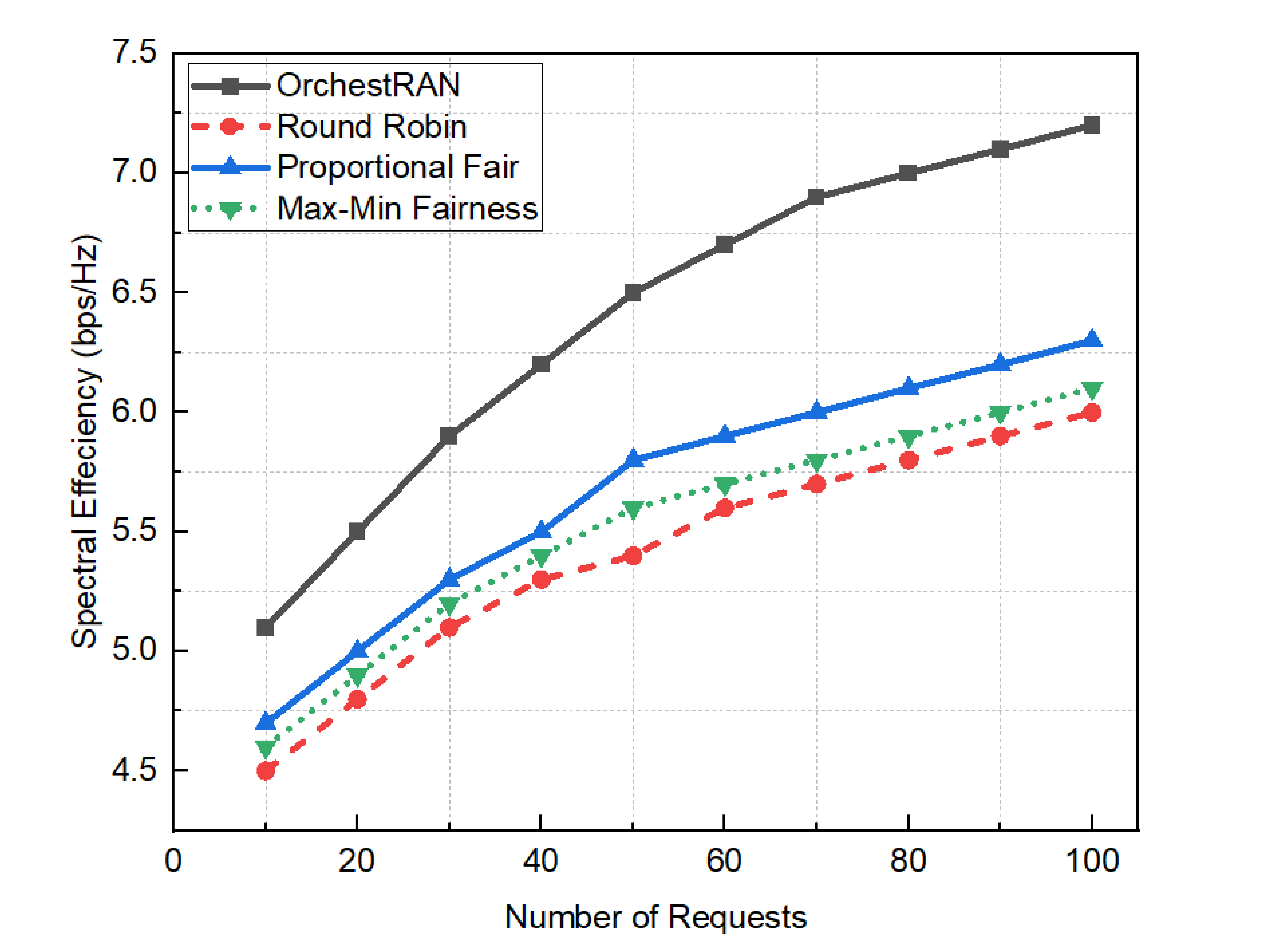}
        \caption{Spectral efficiency comparison}
        \label{fig:subfig22}
    \end{subfigure}
    \caption{Performance comparison of the proposed AI-RAN architecture based on OrchestRAN with traditional scheduling algorithms.}
    \label{fig:orchestrran_architecture_}
\end{figure}

The performance of the proposed framework is analyzed in comparison with traditional scheduling algorithms, including Round Robin, Proportional Fair, and Max-Min Fairness, across different performance metrics such as latency and spectral efficiency. As illustrated in Fig.~\ref{fig:subfig11}, OrchestRAN demonstrates a significantly lower latency compared to other scheduling techniques. The Round Robin algorithm exhibits the highest latency variance, whereas Proportional Fair and Max-Min Fairness provide moderate latency improvements. The reduced latency in OrchestRAN can be attributed to its intelligent decision-making process with AI models based on service requirements. It dynamically allocates resources based on real-time network conditions, and hence minimizing delays.
Further, as illustrated in Fig.~\ref{fig:subfig22}, OrchestRAN consistently outperforms traditional scheduling methods, achieving a spectral efficiency improvement of up to 20\% compared to Round Robin and 10-15\% compared to Proportional Fair and Max-Min Fairness. This superior performance is due to adaptive resource allocation and RL-based optimization, which maximize spectrum utilization while maintaining QoS. Overall, these results validate the effectiveness of the AI-RAN framework based on OrchestRAN, highlighting its capability to enhance network efficiency by reducing latency and improving spectral efficiency under varying network loads.

\subsubsection*{\textbf{Key Takeaways}}
\textit{A new era of network intelligence is driven by transformative technologies such as DTs, GenAI, IRSs, and FL, all of which empower AI-RAN to achieve autonomous, adaptive, and secure operations. Building on this foundation, an advanced AI-RAN architecture has been presented based on the OrchestRAN framework, seamlessly integrating AI-enhanced optimization and dynamic resource allocation mechanisms. Through simulation validation, the proposed architecture demonstrates significant improvements in both latency and spectral efficiency over traditional scheduling methods, establishing its potential to meet—and exceed—the stringent performance demands of future 6G networks.}

\section{Associated Challenges and Future Directions}
Despite the potential of AI-RAN, several challenges hinder its widespread adoption. Besides, the presented concept possesses several future research directions. This section presents these emerging aspects.

\subsection{Technical Challenges}
\label{subsec:technical_challenges}

Designing AI-RAN is challenging due to the diverse requirements of applications, which vary in data type, time granularity, and processing needs. For example, a traffic prediction application may need real-time data from RAN functions, whereas anomaly detection may rely on aggregated metrics from both the RAN and the underlying orchestration or management platform (e.g., the RIC or edge computing infrastructure) over a period of time. Additionally, raw signal measurements might be needed for traffic optimization, while fault detection may depend on analyzing control packet inter-arrival times. Collecting all possible raw data is infeasible due to the high volume and processing overhead. Currently, static Application Programming Interfaces (APIs) defined by bodies like 3GPP expose coarse-grained data suitable for various use cases. However, modifying these APIs requires consensus, making the process slow and limiting flexibility. These constraints hinder the full potential of AI-RAN.

As the architecture of a RAN is disaggregated, deciding the optimal placement of AI application components is challenging. Depending on the location, factors such as latency, network bandwidth, and computational power significantly differ. Orchestrating these AI-RAN applications is particularly challenging. Additionally, AI-RAN applications often share resources with other edge-deployed AI systems (non-RAN) and thus they need to have flexible designs enabling them with proper scalability.

In wireless communication systems, prediction and preparedness for worst-case scenarios are of critical importance to ensure reliable communication. However, the integration of AI into RANs poses challenges due to the non-linear and non-transparent nature of AI models. Even though they perform well in live networks, they struggle to offer worst-case performance guarantees.

Without AI agents, fully autonomous networks would not be possible to achieve. Particularly, even though 2025 has been a banner year for the boom of agentic AI applications, there is limited exploration in the telecommunication sector. This, if not explored further, will hinder the rapid adoption of agents in the telecommunication industry.

Since it is difficult to generate analytical models to verify whether the AI tools are correct, and even more to explain why they work in an intuitive way, AI tools are often addressed as black boxes. The lack of explainability stands in the way of using AI in real-time decision making. So far, cellular networks and wireless standards have historically been designed considering a mixture of theoretical analysis, channel measurements, and human intuition. Therefore, it is important that AI models have the same level of explainability, thereby necessitating further research exploration in the domain of Explainable AI (XAI).

\subsection{Regulatory Challenges}
 \label{subsec:regul_challenges}
6G networks must comply with regulations and ensure security, but embedding AI introduces new concerns. User data used to train AI models may be sensitive. To meet the International Mobile Telecommunications 2030 (IMT-2030) security guidelines of the International Telecommunication Union (ITU), AI must be designed with privacy-by-design principles. Otherwise, models become vulnerable to adversarial attacks, where attackers feed misleading inputs, or poisoning attacks that manipulate training data. 

To address security issues, the research community has explored solutions inspired by cybersecurity. FL is one such method, but it too can be vulnerable, leaking sensitive information through model updates. Due to the vast number of real-time applications and massive data involved, risks such as model memorization arise, where client-shared model parameters reveal indirect insights about private data. Other attack types include inference-phase attacks (e.g., membership inference, model inversion) and training-phase attacks (e.g., sybil, gradient leakage), all of which require attention. Recognizing these risks, 3GPP has introduced AI security enhancements in Release 18, alongside normative work on AI-RAN \cite{3GPP_RP_213602}. Accordingly, implementations must align with standards for interoperability, reliability, and safety. 3GPP TR 33.898 outlines threat models and mitigation strategies to protect AI/ML-based services in RANs \cite{3GPP_TR_33.898}. Security, resilience, and trust are core design targets for 6G in the IMT-2030 framework. AI algorithms used in RANs must meet reliability and safety standards, which is challenging and calls for new testing methods and architectural changes. ITU has issued recommendations (e.g., Y.3172 and related) outlining architectural frameworks for ML in future networks, showing how to integrate learning systems while preserving required determinism.

Traditional rule-based RAN algorithms are predictable and well understood. In contrast, AI models are non-deterministic, making verification, validation, auditing, and regulatory approval more complex. Without XAI-enabled transparency, standardization bodies may hesitate to trust AI for RAN functions. Also, with growing focus on sustainability, AI-RAN must be environmentally friendly. However, AI contributes significantly to carbon emissions, especially during inference. Therefore, developing energy-efficient AI models and optimizing data center operations are essential for compliance.

\subsection{Future Prospects}
As wireless networks approach 6G and beyond, AI will be at the core of the design and operation of RANs, thus transforming the static infrastructure into an intelligent and adaptive infrastructure capable of autonomous operation. In regard to this, the following discussions present key forward-looking research ideas.%

\subsubsection{RAN as a Multi-Agent System (RAN-MAS)} 
In future architectures, we envision the RAN evolving into a Multi-Agent System (MAS), wherein each RAN node functions as an AI agent capable of perception, reasoning and autonomous decision-making. This enables the network to dynamically adjust spectrum usage, handover policies, and power levels anticipating network conditions. Furthermore, RAN-MAS introduces the potential for multi-agent collaboration, wherein RAN nodes intelligently coordinate to perform tasks such as interference management, CoMP transmission, and network slicing orchestration.
\subsubsection{AI-Empowered Integrated Sensing and Communication (ISAC)} The amalgamation of communication and sensing capabilities is expected to be a fundamental aspect in 6G and beyond. This can enable a network to transmit data and sense the environment simultaneously. This duality can allow base stations to perform tasks such as object detection, localization, environmental mapping, and various others alongside communication, thus contributing to various applications including, but not limited to, vehicular networks, environmental monitoring, and smart city development. Currently, research is being conducted on the synergy of ISAC with IRSs and other key enablers, among which AI empowerment for ISAC and its integration into RAN architectures are active research areas. These trends unlock insights into how AI-RAN is expected to use novel AI/ML algorithms and approaches like FL to empower ISAC. Alongside this, continual research on implementation requirements of ISAC such as waveform design, optimization, and detection are also necessary. 

\subsubsection{Energy-Efficient AI} 
As energy demand grows with the complexity of AI, AI-RAN is expected to balance intelligence with sustainability. To this end, research on light weight AI models and neuromorphic accelerators must be adopted, thereby reducing power consumption.

\subsubsection{Decentralized Autonomous Organizations (DAOs) for Intelligent RAN Orchestration} 
DAOs can facilitate intelligent RAN orchestration by autonomously managing AI model deployment and policy enforcement. This enhances trust, scalability, and coordination across distributed RAN nodes.

\subsubsection*{\textbf{Key Takeaways}}
\textit{The previous sections summarize that the perspective RAN brings both the challenges and future research directions. For instance, AI-RAN faces critical challenges in data handling, stemming from the need to process heterogeneous inputs and manage complex orchestration across disaggregated nodes. On the regulatory front, concerns such as preserving FL privacy, and mitigating threats demand the adoption of standardized, low-power AI models and strict adherence to architecture-compliant integration aligned with 3GPP and ITU standards. Looking ahead, the future of AI-RAN envisions the adoption of multi-agent systems for fully autonomous RAN operations, the integration of technologies, the development of ultra energy-efficient AI through lightweight models.}

\section{Conclusion}
This study has highlighted the potential of integrating AI into RAN architectures, thereby enabling intelligent, self-optimizing, and fully automated networks for 6G and beyond. The proposed AI-RAN framework demonstrates improved network performance and adaptability. However, challenges such as explainability, scalability, data protection, and compliance must be addressed. Further research is needed, particularly in security protocols, standardized AI agent architectures, and sustainable AI operations, to fully realize AI-RAN.

\section*{Acknowledgment}
This work was supported in part by the Senate Research Committee (SRC) of the University of Moratuwa, Sri Lanka, under Grant SRC/ST/2024/23, and in part by the U.S National Science Foundation under Grant ECCS-2335876.

\ifCLASSOPTIONcaptionsoff
  \newpage
\fi










\begin{thebibliography}{1}

\bibitem{gong2024holographic}
T. Gong, P. Gavriilidis, R. Ji, C. Huang, G. C. Alexandropoulos, L. Wei, Z. Zhang, M. Debbah, H. V. Poor, and C. Yuen, 
``Holographic MIMO Communications: Theoretical Foundations, Enabling Technologies, and Future Directions,'' 
\textit{IEEE Communications Surveys \& Tutorials}, vol. 26, no. 1, pp. 196--257, 2024, doi: 10.1109/COMST.2023.3309529.

\bibitem{10731639}
N. Sehad, L. Bariah, W. Hamidouche, H. Hellaoui, R. Jantti, and M. Debbah, 
``Generative AI for Immersive Communication: The Next Frontier in Internet-of-Senses Through 6G,'' 
\emph{IEEE Communications Magazine}, vol.~TBD, no.~TBD, pp.~1--13, 2024, doi: 10.1109/MCOM.001.2400199.

\bibitem{frauendorf2023evolution}
J. L. Frauendorf and É. Almeida de Souza, 
``The Evolution of RAN (Radio Access Network), D-RAN, C-RAN, V-RAN, and O-RAN,'' 
in \textit{The Architectural and Technological Revolution of 5G}, Springer, Cham, 2023. DOI: \url{https://doi.org/10.1007/978-3-031-10650-7_10}.

\bibitem{Chih-Lin2014Recent}
I. Chih-Lin, J. Huang, R. Duan, C. Cui, J. X. Jiang, and L. Li, 
``Recent Progress on C-RAN Centralization and Cloudification,'' 
\emph{IEEE Access}, vol. 2, pp. 1030--1039, 2014, doi: 10.1109/ACCESS.2014.2351411.

\bibitem{article1}
N. A. Kaim Khani and S. Schmid, 
``AI-RAN in 6G Networks: State-of-the-Art and Challenges,'' 
\emph{IEEE Open Journal of the Communications Society}, vol. PP, pp. 1--1, Jan. 2023, doi: 10.1109/OJCOMS.2023.3343069.

\bibitem{9711524}
L. U. Khan, W. Saad, D. Niyato, Z. Han, and C. S. Hong, 
``Digital-Twin-Enabled 6G: Vision, Architectural Trends, and Future Directions,'' 
\emph{IEEE Communications Magazine}, vol. 60, no. 1, pp. 74--80, 2022, doi: 10.1109/MCOM.001.21143.

\bibitem{10685369}
H. Zhou, C. Hu, Y. Yuan, Y. Cui, Y. Jin, C. Chen, H. Wu, D. Yuan, L. Jiang, D. Wu, X. Liu, C. Zhang, X. Wang, and J. Liu, 
``Large Language Model (LLM) for Telecommunications: A Comprehensive Survey on Principles, Key Techniques, and Opportunities,'' 
\emph{IEEE Communications Surveys \& Tutorials}, vol.~TBD, no.~TBD, pp.~1--1, 2024, doi: 10.1109/COMST.2024.3465447.

\bibitem{zhang2024xlamfamilylargeaction}
J. Zhang, T. Lan, M. Zhu, Z. Liu, T. Hoang, S. Kokane, W. Yao, J. Tan, A. Prabhakar, H. Chen, Z. Liu, Y. Feng, T. Awalgaonkar, R. Murthy, E. Hu, Z. Chen, R. Xu, J. C. Niebles, S. Heinecke, H. Wang, S. Savarese, and C. Xiong, 
``xLAM: A Family of Large Action Models to Empower AI Agent Systems,'' 
\emph{arXiv Preprint}, arXiv:2409.03215, 2024. [Online]. Available: \url{https://arxiv.org/abs/2409.03215}.

\bibitem{nakamoto2008}
S. Nakamoto, 
``Bitcoin: A Peer-to-Peer Electronic Cash System,'' 
2008. [Online]. Available: \url{https://bitcoin.org/bitcoin.pdf}. [Accessed: Feb. 27, 2025].

\bibitem{9178307}
L. Bariah, L. Mohjazi, S. Muhaidat, P. C. Sofotasios, G. K. Kurt, H. Yanikomeroglu, and O. A. Dobre, 
``A Prospective Look: Key Enabling Technologies, Applications and Open Research Topics in 6G Networks,'' 
\textit{IEEE Access}, vol. 8, pp. 174792--174820, 2020, doi: 10.1109/ACCESS.2020.3019590.

\bibitem{10199530}
J. Guo, L. Gao, N. Li, S. Yang, J. Zhu, X. She, J. Wang, and P. Chen, 
``System-Level Simulation and Performance Evaluation for 6G Ultra Massive MIMO,'' 
in \emph{Proc. 2023 IEEE 97th Vehicular Technology Conference (VTC2023-Spring)}, 
2023, pp. 1--6, doi: 10.1109/VTC2023-Spring57618.2023.10199530.

\bibitem{10379539}
Z. Wang, J. Zhang, H. Du, D. Niyato, S. Cui, B. Ai, M. Debbah, K. B. Letaief, and H. V. Poor, 
``A Tutorial on Extremely Large-Scale MIMO for 6G: Fundamentals, Signal Processing, and Applications,'' 
\emph{IEEE Communications Surveys \& Tutorials}, vol. 26, no. 3, pp. 1560--1605, 2024, 
doi: 10.1109/COMST.2023.3349276.

\bibitem{10786352}
C. Sandeepa, E. Zeydan, T. Samarasinghe, and M. Liyanage, 
``Federated Learning for 6G Networks: Navigating Privacy Benefits and Challenges,'' 
\emph{IEEE Open Journal of the Communications Society}, vol. 6, pp. 90--129, 2025, 
doi: 10.1109/OJCOMS.2024.3513832.

\bibitem{3GPP_RP_213602}
3GPP, 
``Artificial Intelligence (AI)/Machine Learning (ML) for NG-RAN,'' 
3GPP TSG RAN Meeting \#94e, RP-213602, Dec. 2021. [Online]. Available: \url{https://www.3gpp.org/ftp/tsg_ran/TSG_RAN/TSGR_94e/Docs/RP-213602.zip}

\bibitem{3GPP_TR_33.898}
3GPP, 
``Study on Security and Privacy of Artificial Intelligence/Machine Learning (AI/ML)-Based Services and Applications in 5G,'' 
3GPP TR 33.898, June 2023. [Online]. Available: \url{https://www.3gpp.org/dynareport/33898.htm}

\end{thebibliography}
\end{document}